\documentclass[12pt,a4paper]{article}

\usepackage{amsmath,amsthm}
\usepackage{amssymb,latexsym}
\usepackage{amsfonts}
\newtheorem{theorem}{Theorem}[section]

\newtheorem{lemma}{Lemma}[section]

\newtheorem{remark}{Remark}[section]
\newtheorem{assumption}{Assumption}[section]
\numberwithin{equation}{subsection}
\usepackage{enumerate}
\begin{document}
\thispagestyle{empty}
{\bf \Large  Group Signature Schemes Using Braid Groups}
\begin{center}
Tony Thomas, Arbind Kumar Lal\\
Department of Mathematics and Statistics\\
Indian Institute of Technology Kanpur, Kanpur\\
Uttar Pradesh, India-208016\\
\{tony,arlal\}@iitk.ac.in
\end{center}
\begin{abstract}
Artin's braid groups have been recently suggested as a new source for public-key cryptography. In this paper we propose the first group signature schemes based on the conjugacy problem, decomposition problem and root problem in the braid groups which are believed to be  hard problems.
\end{abstract}
\section{Introduction}
Braid groups have recently attracted the attention of many cryptographers as an alternative to number-theoretic public-key cryptography. The birthdate of braid group based cryptography can be traced back to the pioneering work of Anshel {\it et al.} in 1999~\cite{Anshel2:etal} and Ko {\it et al.} in 2000~\cite{Ko:etal:2000}. Since then, braid groups have attracted the attention of many cryptographers due to the fact that, they provide a rich collection of hard problems like the {\itshape conjugacy problem, braid decomposition problem} and {\itshape root problem}  and there are efficient algorithms for parameter generation and group operation~\cite{Cha:etal}.\par
 Since the construction of a Diffie-Hellman type key agreement
protocol and a public key encryption scheme by Ko {\it et al.} in
2000~\cite{Ko:etal:2000}, there have been many attempts to design
other cryptographic protocols using braid groups. Positive results
in this direction are a construction of pseudorandom number generator by Lee {\it et al.} in 2001~\cite{Lee:etal}, key agreement protocols by Anshel {\it et al.}
in 2001~\cite{Anshel:etal}, an implementation of braid
computations by Cha {\it et al.} in 2001~\cite{Cha:etal},
digital signature schemes by Ko {\it et al.} in 2002~\cite{Ko:etal:2002}, entity authentication schemes by Sibert {\it et al.} in 2002~\cite{Sibert:etal} and a provably-secure identification scheme by Kim {\it et al.} in 2004~\cite{Kim}.\par
Digital signatures bind signers to the contents of the document they sign. Group signature schemes were introduced by Chaum and van Heyst~\cite{Chaum:91} to allow individual members of a group to sign messages on behalf of a group. Formally a group signature scheme has the following properties~\cite{Chaum:91}:
\begin{enumerate}
\item only members of the group can sign messages;
\item the receiver of the signature can verify that it is a valid signature of the group, but cannot identify the signer;
\item in case of a dispute at a later stage, the signature can be opened to reveal the identity of the signer.
\end{enumerate}
The salient features of group signatures make them attractive for many specialized applications, such as voting and bidding. More generally, group signatures can be used to conceal organizational structures, e.g., when a company or a government agency issues a signed statement. Group signatures can also be integrated with an electronic cash system whereby several banks can securely distribute anonymous and untraceable e-cash. \par
\indent Group signatures are generalization of {\it credential mechanisms} (\cite{dch}) and of {\it membership authentication schemes} (~\cite{ook},~\cite{shi}), in which a group member can convince a verifier that he belongs to a certain group without revealing his identity.\par
In this paper, we design  some  group signature schemes using
braid groups. These are the first group signature schemes using
braid groups.\par 
In Section 2, we briefly review the
basics of braid groups. We describe the initial system set up
and some security assumptions needed for building up these
signature schemes in Section 3. A group signature scheme whose security is
based on the root problem is described in Section 4. In Section
5, we describe a group signature scheme that employ confirmation and denial
protocols for identifying the actual signer. The security of this
scheme is based on the root problem, conjugacy problem and its
variants. A third group signature scheme whose security is based
on the conjugacy problem and its variants is described in Section
6. The paper concludes with some general remarks in Section 7.
\section{An Overview of Braid Groups}
In this section, we briefly describe the basics of braid groups, hard problems in braid groups. A good introduction to braid groups is~\cite{Birman} and survey articles on braid cryptography are~\cite{Eonkyung},~\cite{Deho}.
\subsection{Geometric Interpretation of Braids}
A braid group $B_{n}$ is an infinite non-commutative group arising from geometric braids composed of $n$-strands.
 A braid is obtained by laying down a number of parallel strands and intertwining them so that they run in the same direction. The  number of strands is called the braid index. Braids have the following geometric interpretation: an $n$-braid (where $n \in \mathbb{N}$) is a set of disjoint $n$ strands all of which are attached to two horizontal bars at the top and bottom such that each strand always heads downwards as one moves along the strand from top to bottom. Two braids are equivalent if one can be deformed to the other continuously in the set of braids.\par
 Let $B_n$ be the set of all $n$-braids. $B_{n}$ has a natural group structure. Each $B_{n}$ is an infinite torsion-free noncommutative group and its elements are called $n$-braids.  The multiplication {\itshape ab} of two braids {\itshape a} and {\itshape b} is the braid obtained by positioning {\itshape a} on the top of {\itshape b}. The identity {\itshape e} is the braid consisting of $n$ straight vertical strands and the inverse of {\itshape a} is the reflection of {\itshape a} with respect to a horizontal line. \par
Let $\mathbf{S}_n$ be the symmetric group on $n$ symbols. Given a braid $\alpha$, the strands define a map $p(\alpha)$ from the top set of endpoints to the bottom set of endpoints. In this way we get a homomorphism  $p : B_n \rightarrow S_n$.
\subsection{Presentations of Braid Groups}
 Any braid can be decomposed as a product of simple braids known as {\itshape Artin generators} $\sigma_{i}$, that have a single crossing between the $i^{th}$ strand and the $(i+1)^{th}$ strand with the convention that the $i^{th}$ strand crosses under the $(i+1)^{th}$ strand. The homomorphism,  $p$ maps the generator $\sigma_i$  to the transposition $\tau_i$ $(=(i,i+1))$. \par
For each integer $n \geq\ 2$, the $n$-braid group $B_{n}$ has the  Artin presentation by generators $ \sigma_{1},\sigma_{2},\dots,\sigma_{n-1}$ with relations
\begin{equation}
\begin{split}
   &\ \sigma_{i}\sigma_{j} = \sigma_{j}\sigma_{i}, ~\mbox{where}~~|i-j|\geq 2,~~\mbox{and} \\
   &\ \sigma_{i}\sigma_{i+1}\sigma_{i} = \sigma_{i+1}\sigma_{i}\sigma_{i+1},~\mbox{for}~~ 1 \leq i \leq n-2.
\end{split}
\end{equation}
\subsection{Some Special Classes of Braids}
  Let $B_{n}^{+}$ denote the submonoid of $B_{n}$ generated by $\{ \sigma_{1},\dots,\sigma_{n-1} \}$. Elements of $B_{n}^{+}$ are called the {\itshape positive braids}. A positive braid is characterized by the fact that at each crossing the string going from left to right undercrosses the string going from right to left. \par A positive braid is called {\it non-repeating}
if any two of its strands cross at most once. We denote $D = D_n \subset B_{n}^{+}$ to be the set of all non-repeating braids. To each $\pi \in S_n$ we can
associate a unique $\alpha \in D_n$ in the following way : for $i =
1,\dots,n$ connect the upper $i$-th point to the lower $\pi(i)$-th
point by a straight line making each {\it crossing positive}, {\it
i.e.} the line between $i$ and $\pi(i)$ is under the line between
$j$ and $\pi(j)$ if $ i < j$. The following lemma says that $p$
restricted to $D_n$ is a bijection.
\begin{lemma}~\cite{Epstein:etal}~\label{lemma:geodesic}
The homomorphism $p : B_n \rightarrow S_n$ restricted to $D_n$ is a bijection.
\end{lemma}
Hence non-repeating braids are also known as {\it permutation braids}. From this lemma it follows that $|D_n| = n!$. In this way we can identify $D_n$ with $S_{n}$ . \par
Let $LB_{n}$ and $RB_{n}$ be two subgroups of $B_n$  consisting of braids obtained by braiding left $\lfloor \frac{n}{2} \rfloor$ strands and right $n-\lfloor  \frac{n}{2} \rfloor$ strands, respectively. That is,
\[
LB_n = \langle \sigma_1, \dots,\sigma_{\lfloor \frac{n}{2} \rfloor -1}\rangle, \mbox{ and } RB_n = \langle \sigma_{\lfloor \frac{n}{2} \rfloor +1}, \dots, \sigma_{n-1} \rangle.
\]
Then we have the commutativity property that for any $\alpha \in LB_{n}$ and $\beta \in RB_{n}$, $\alpha \beta = \beta \alpha$. These subgroups of $B_n$ are used in designing various cryptographic protocols.
\subsection{Canonical Decomposition of Braids}
For two words $v$ and $w$ in $B_{n}$, we say that $v \leq w$, if $w = avb$ for some $a, b \in B_{n}^{+}$. Then $\leq$ is a partial order in $B_{n}$~\cite{Epstein:etal}.\par
The positive braid, $\Delta =
(\sigma_1\dots\sigma_{n-1})(\sigma_1\dots\sigma_{n-2})\dots(\sigma_1\sigma_2)\sigma_{1}$ is called the {\it fundamental braid}. A braid satisfying $e \leq A
\leq \Delta$ is called a {\itshape canonical factor}. There is a bijection between the set of all permutation braids and the set of all canonical factors~\cite{Epstein:etal}. Thus a canonical factor can be denoted by the corresponding permutation $\pi \in S_{n}$. By $\pi_{\Delta}$, we mean the permutation corresponding to the fundamental braid $\Delta$.\par
  For a positive braid $P$, we say that the decomposition $P = A_0P_0$ is {\it left-weighted} if $A_0$ is a canonical factor, $P_0 \geq e$ and $A_0$ has the maximal word length among all such decompositions. A left-weighted decomposition  $P = A_0P_0$ is unique~\cite{Cha:etal}. $A_0$ is called the {\it maximal head} of $P$. Any braid $x$ can be uniquely decomposed as
\begin{equation}
x = \Delta^{u}A_1A_2\dots A_k,~~\mbox{where}~~u \in \mathbb{Z}, A_i \neq e,\Delta,~~\mbox{is a canonical factor}
\end{equation}
and the decomposition $A_iA_{i+1}$ is left-weighted for each $1 \leq i \leq k-1$~\cite{Cha:etal}. This unique decomposition is called the {\it left canonical form} of $x$ and so it solves the word problem. Since each canonical factor corresponds to a permutation braid, $x$ can be denoted as
\begin{equation}
x = \pi_{f}^{u}\pi_1\pi_2\dots \pi_k,~~\mbox{where}~~\pi_i \neq Identity, \pi_{f}.
\end{equation}
Hence for implementation purposes the braid $x$ can be represented as the tuple $(u, \pi_1, \pi_2, \dots, \pi_k)$. The integer $u$, denoted by $\inf(x)$ is called the {\it infimum} of $x$ and the integer $u+k$, denoted by $\sup(x)$ is called the {\it supremum} of $x$. The {\it canonical length} of $x$, denoted by len(x), is given by $k = \sup(x) - \inf(x)$.\par
\subsection{Hard Problems in Braid Groups}
We use the following hard problems in our signature schemes.
\begin{enumerate}
\item{\bf Conjugacy Search Problem (CSP)}\\
Let $(x,y) \in B_{n}\times B_n$, such that $y=a^{-1}xa$, where $a \in B_{n}$ or some subgroup of $B_n$. The {\itshape conjugacy search problem} is to find  a $b$ such that $y = b^{-1}xb$.
\item{\bf Multiple Simultaneous Conjugacy Search Problem (MSCSP)}\\
Let $(x_1, a^{-1}x_1a),\dots,(x_r, a^{-1}x_ra) \in B_n \times B_n $ for some $a \in B_n$ or some subgroup of $B_n$. The {\itshape multiple simultaneous conjugacy problem} is to find a $b$ such that, $ b^{-1}x_1b = a^{-1}x_1a,~ \dots,~ b^{-1}x_rb = a^{-1}x_ra.$
\item{\bf Braid Decomposition Problem (BDP)}\\
Let $(x, y) \in B_{n}\times B_n$, where $y = a_{1}xa_{2}$ for some $(a_{1}, a_{2}) \in LB_{n}\times LB_n$. The {\it braid decomposition problem} is to find a pair $(b_{1}, b_{2})\in LB_{n}\times LB_n$ such that $y = b_{1}xb_{2}$.
\item{\bf Multiple Simultaneous Braid Decomposition Problem (MSBDP)}\\
Let $(x_1, a_{1}x_1a_{2}), \dots,(x_r, a_{1}x_ra_{2}) \in B_n \times B_n$ for some $(a_1,a_2) \in LB_n\times LB_n$. The {\itshape multiple simultaneous braid decomposition problem} is to find a pair $(b_1,b_2) \in LB_n\times LB_n $ such that, $b_{1}x_1b_2 = a_{1}x_1a_{2},~ \dots,~ b_{1}x_rb_{2} = a_{1}x_ra_{2}.$
\item{\bf Root Extraction Problem (RP)}\\
Let $x = a^p$, where $a, x \in B_n$ and $p \in \mathbb{N}$. Then the {\itshape root problem} (for the exponent $p$) is to find a braid $b \in B_n$ such that $b^{p} = x$.\par
\end{enumerate}
\section{Preliminaries}
In this section, the initial system set up, intractability assumptions, some other assumptions and some notation used in this paper are given.
\subsection{Initial Setup}
The system parameters $n$ and $l$ are chosen to be sufficiently large positive integers and are made public. Since the braid group $B_{n}$ is discrete and infinite, we cannot have a uniform  probability distribution on $B_{n}$. But there are finitely many positive $n$-braids with $l$ canonical factors, we may consider randomness for these braids. Such a braid can be generated by concatenating $l$ random canonical factors. Let,
\begin{eqnarray*}
B_{n}(l) &= &\{ b \in B_{n} \ \vert \ 0 \leq inf(b) \leq sup(b) \leq l \},\\
LB_{n}(l) &=& \{ b \in LB_{n} \ \vert \ 0 \leq \inf(b) \leq \sup(b) \leq l \}~~\text{and}
\end{eqnarray*}
\begin{eqnarray*}
RB_{n}(l) &=& \{ b \in RB_{n} \ \vert \ 0 \leq \inf(b) \leq \sup(b) \leq l \}.
\end{eqnarray*}
Then $\vert B_{n}(l)\vert \leq l(n!)^{l}$ and so $LB_{n}(l), RB_{n}(l)$ and $B_{n}(l)$ are finite sets. We use the random braid generator given in ~\cite{Cha:etal} (which produces random braids in $O(ln)$ time) for generating random braids. Also, we consider uniform probability distribution on these sets. \par
Let $H : \{0,1\}^{*} \rightarrow B_{n}(l)$  be a collision free hash function. $H$ can be constructed by composing a usual hash function of bit strings with a conversion from bit strings of fixed length
to elements of $B_n(l)$. A way to construct this conversion function,
$c : \{0,1\}^{k}\rightarrow B_n(l)$ is given in~\cite{Ko:etal:2002}.
\subsection{Notations}
We use the following notations through out this paper.
\begin{itemize}
\item By $a \in_r A$, we mean a random choice of an element $a$ from the set $A$.
\item By $ P \stackrel{Q}{\longrightarrow} V$, we mean $P$ sends $Q$ to $V$.
\end{itemize}
\subsection{Group Manager}
Let $T$ be a group manager, who chooses the private key of the group and creates the public key of the group. $T$ also manages the members of the group. $T$ is needed in identifying the actual signer in our first and third signature schemes. $T$ is not needed in our second signature scheme.
\subsection{Intractability Assumptions}
We assume that the hard problems {\bf CSP, MSCSP, BDP, MSBDP, RP}, stated in Section 2.5 are intractable in braid groups. However, we assume that the {\it conjugacy decision problem} given below  is easy in braid groups.\par
Let $(x,y) \in B_{n}\times B_n$. The {\itshape conjugacy decision problem} is to decide whether $x$ and $y$ are conjugates or not, that is to decide whether there exists an $a \in B_n$ such that $y = a^{-1}xa$ or not. The conjugacy decision problem may be solved using the algorithm given in~\cite{Ko:etal:2002}.
\subsection{Some New Assumptions}
In this paper, we make two assumptions. The first assumption is similar to the {\bf EDL} intractability
assumption used in~\cite{Lyu}. The {\bf EDL} (Equality of
Discrete Logarithms) intractability assumption can be stated as
follows : given $x, y \in_r G = \langle f \rangle =\langle g
\rangle$, it is computationally infeasible to determine the
equality of $\log_{f}x$ and  $\log_{g}y$ over $\mathbf{Z}_n$,
where $ord(g)=n$. So we have our first assumption as
\begin{assumption}
For $(\alpha, \beta) \in B_n \times B_n$, let \[F_{\beta}(\alpha) = \{(a, b) \in B_n \times B_n: \alpha= a\beta b\}.\]
Then, given two pairs of braids $(\alpha, \beta)$ and $(\gamma, \delta)$ in $B_n \times B_n$, it is computationally infeasible to check whether $F_{\beta}(\alpha) \cap F_{\delta}(\gamma) \neq \emptyset$ or not.
\end{assumption}
The second assumption is about cardinalities of certain sets, which may be stated as follows.
\begin{assumption}
Let $n$, $l$ be sufficiently large positive integers, $\alpha, \beta, \gamma \in_r B_{n}(l)$, $a_1,a_2  \in_r LB_{n}(l)$ and $a  \in_r RB_{n}(l)$. Then the cardinality of the set\[E_{a}(\beta,\gamma) = \{b \in RB_{n}(l):~ b^{-1}\alpha b= a^{-1}\alpha a, ~b^{-1}\beta b \neq a^{-1}\beta a\}\]
 is bounded below by a non decreasing function $p(n,l)$ of $n$ and $l$.
\end{assumption}
In this paper, we do not undertake any theoretical or numerical study to check the validity of the above assumptions. 
\section{Group Signature Scheme 1}
In~\cite{Chaum:91} Chaum {\it et al.} describe a group signature
scheme using public-key systems. In this case the group manager
$T$ chooses a public key system, gives each person a list of
secret keys (these lists are all disjunct) and publishes the
complete list of corresponding public keys (in random order) in a
Trusted Public Directory. Each person can sign a message with a
secret key from his list, and the recipient can verify this
signature with the corresponding public key from the public list.
Each key will be used only once, otherwise the signature created
with that key gets linked. $T$ knows all the list of secret keys,
so that in case of a dispute, he can identify the signer. Hence
$T$ is needed for the setup and for opening of the signature.\par
We can adopt this group signature scheme directly to the braid
group frame work as follows : $T$ chooses a set $E$ of braids and
raises them to the $p^{th}$ power, where $p$ is an integer greater
than $1$. Each person is given a list of braids from $E$  (these
lists are all disjunct) and the complete list of $p^{th}$ powers
of elements of $E$ (in random order) is published in a Trusted
Public Directory. To sign a message $m$, a group member chooses a
braid $\alpha$ from his list and forms the signature $S_m = \alpha
H(m)$. The recipient can verify this signature by computing $(S_m
H(m)^{-1})^p$ and checking it with the corresponding public key in
the Trusted Public Directory. Each key will be used only once. $T$
knows all the list of secret keys, so that in case of a dispute,
he can identify the signer.\par 
A problem with this scheme is that
the group manager knows all the secret keys of the group members
and can therefore also create signatures. This problem can be overcome by making each user to untraceably send one (or more) public keys to a public list,
which will be the public key of the group. But it has to be ensured that only the group members will be able to send public keys to that list.\par
Although, the scheme is very elegant it has the obvious
disadvantage that a key can be used only once. However, we can
trivially see that the security of this scheme is equivalent to
solving the root problem. Hence this group signature scheme is
highly secure. This is the only cryptographic scheme on braid
groups whose security depends solely on the root problem (RP). 
\section{Group Signature Scheme 2}
In this section, we describe a group signature scheme which does not involve a group manager. The security of the scheme is based on the hardness of {\bf BDP, MSCSP} and {\bf RP}. Here the recipient of the signature can easily check whether the signature has come from a particular group or not. But the identity of the signer can not be verified unless the verifier engages in an interactive protocol with the signer as in the case of undeniable signatures.\par
\subsection{Key Generation}
 Let $G$ be a group with $k$ members $P_{1},P_{2},\dots,P_{k}$. The members of the group agree on a secret braid $\alpha \in B_{n}(l)$. $\beta = \alpha^{4}$ is published as the public key of the group. Also, each member $P_{i}$ of the group chooses $(u_{i},v_{i}) \in LB_{n}(l) \times LB_{n}(l)$ as his secret key. In this case, the public key of $P_{i}$ is $x_{i} = u_{i}^{-1}\beta v_{i}$.\par
We shall denote by $PK$ the tuples $(\beta, \{x_i\}_{1}^{k})$ generated as above.
\subsection{Signature Generation}
Let $m$ be the message to be signed. Suppose $P_{i}$ wants to sign $m$. He computes the signature $S_m = u_{i}^{-1}y^{-1}\alpha^{2}yu_{i}$, where  $y = H(m)$.\par
We shall denote by $SIG(m)$, the set of valid signatures on $m$.
\subsection{Confirming the Group Identity of the Signature}
Given an alleged signature $\hat{S}_m$, suppose that a verifier $V$ wants to check whether it is a valid signature from the group $G$. $V$ computes $\hat{S}_m^{2}$ and checks whether it is conjugate to $\beta$ using the algorithm described in~\cite{Ko:etal:2002}.\par
Note that $\hat{S}_m^{2} = u_{i}^{-1}y^{-1}\beta yu_{i}$. Hence if  $\hat{S}_m$ is a valid signature of a member of $G$, then $S_m^{2}$ is conjugate to $\beta$.
\subsection{Confirmation Protocol}
Suppose that a signer $P_i$ claims that a signature $\hat{S}_m$ was made by him. Then a verifier $V$ first checks the group identity of the signature using the above protocol and then verifies the claim of $P_i$ by engaging in an interactive confirmation protocol with him. Let us denote the prover $P_{i}$ by $P$. When $\hat{S}_m$ is a valid signature of $m$ by $P$, he will be able to convince $V$ of this fact, while if the signature is invalid then no prover even if he is computationally unbounded will be able to convince $V$ to the contrary except with a negligible probability.\\\\\\\\
{\bf Signature Confirmation Protocol}\\\\
Input : Prover: Secret keys  ($\alpha, u_{i}, v_{i}$). \\
\hspace*{1.2cm} Verifier: Public key  $(\beta,\{x_{j}\}_{j=1}^{k})$ and alleged $\hat{S}_m$.
\begin{enumerate}
\item $V$ chooses $a \in_r RB_{n}(l)$, computes $Q = a^{-1}(\hat{S}_{m})^{2}x_{i}a$ and $ V \stackrel{Q}{\rightarrow} P$.
\item $P$ chooses  $b,c \in_r B_{n}(l)$, computes $R = bu_{i}Qv_{i}^{-1}c $ and $P \stackrel{R}{\rightarrow} V$.
\item $V \stackrel{a}{\rightarrow} P$.
\item $P$ Checks the value of $Q$ and then  $P \stackrel{(b,c)}{\rightarrow} V$.
\item $V$ verifies that $ R  = ba^{-1}y^{-1}\beta y\beta ac$.
\end{enumerate}
If equality holds then $V$ accepts $\hat{S}_m$ as the signature on $m$, otherwise ``undetermined".
\begin{theorem} {\bf Confirmation Theorem.} Let $(\beta, \{x_i\}_{1}^{k}) \in PK$.\\
{\bf Completeness}: Given $S_m \in SIG(m)$, if $P$ follows the {\itshape signature confirmation protocol} then $V$ always accepts $S_m$ as a valid signature.\\
{\bf Soundness}: A Cheating prover $P^{*}$ even computationally unbounded cannot convince $V$ to accept $\hat{S}_m \notin SIG(m)$ with probability greater than $\frac{1}{p(n,l)}$.
\end{theorem}
\begin{proof}
{\bf Completeness}: Let $S_m$ be a valid signature. $P$ computes
\begin{equation*}
R = b(u_iQv_{i}^{-1})c = b(u_{i}a^{-1}(\hat{S}_m)^{2}x_{i}av_{i}^{-1})c =  ba^{-1}(y^{-1}\beta y \beta)ac.
\end{equation*}
which $V$ verifies after getting $(b,c)$ from $P$ and accepts the signature as valid. Hence the protocol is complete.\\
{\bf Soundness}: The idea is that there are many values of $a$ which give the same value for the challenge $Q$ and different values for the response $R$  and a cheating prover $P^*$ has no way to distinguish between these different values of $a$, even if he has infinite computational power. That is, from Assumption 3.2, there are at least $p(n,l)$ choices, for $a \in RB_{n}(l)$ which give the same value of $Q$  but giving different values of $R$. Hence it is infeasible for a cheating prover $P^*$ to distinguish between these different values of $a$, even if he has infinite computational power. Therefore a cheating prover $P^{*}$, even computationally unbounded, cannot convince $V$ to accept $\hat{S}_{m}\notin  SIG(m)$ with probability greater than $\frac{1}{p(n,l)}$. Thus the protocol is sound.
\end{proof}
\begin{remark}
A closer examination of the protocol reveals that it has the zero-knowledgeness property also (see\cite{tony}).
\end{remark}
\subsection{Disavowal Protocol}
If $P_i$ wants to prove to $V$ that $\hat{S}_m$ is not his signature on $m$, he engages in a disavowal protocol with $V$. As in the case of confirmation protocol, we denote $P_i$ by $P$. In the case that $\hat{S}_m$ is not a valid signature, $P$ will be able to convince $V$ of this  fact, while if $\hat{S}_m$ is a valid signature of $P$ on $m$, even if he is computationally unbounded he will not be able to convince $V$ that the signature is invalid except with negligible probability.\\\\
{\bf Disavowal Protocol}\\\\
Input : Prover : Secret keys ($\alpha, u_{i}, v_{i}$). \\
\hspace*{1.2cm} Verifier : Public key  $(\beta,\{x_{j}\}_{1}^{k}) \in PK, y$ and alleged $\hat{S}_m$.
\begin{enumerate}
\item $V$ chooses  $a, b \in_r RB_{n}(l)$ such that $ a$ and $b$ commute and computes\\ $ Q_{1} = a^{-1}(\hat{S}_m)^{2}x_{i}a$, $Q_{2} = b^{-1}(\hat{S}_m)^{2}x_{i}b $ and $ V \stackrel{(Q_{1},Q_2)}{\longrightarrow} P$.
\item $P$ computes the response $R_{1} = u_{i}Q_{1}v_{i}^{-1}$, $R_{2} = u_{i}Q_{2}v_{i}^{-1}$ and $P \stackrel{(R_{1},R_2)}{\longrightarrow} V$.
\item $V$ verifies that $b^{-1}(R_{1}\beta^{-1})b = a^{-1}(R_{2}\beta^{-1})a$.
\end{enumerate}
If equality holds $V$ accepts $\hat{S}_m$ as an invalid signature. Otherwise $P$ is answering improperly.
\begin{theorem} {\bf Denial Theorem} Let $(\beta, \{x_i\}_{1}^{k}) \in PK$.\\
{\bf Completeness}: Suppose that $\hat{S}_m \notin SIG(m)$. If $P$ and $V$ follow the protocol, then $V$ always accepts that $\hat{S}_m$ is not a valid signature of $m$.\\
{\bf Soundness}: Suppose that $\hat{S}_m \in SIG(m)$. Then a cheating prover, even computationally unbounded, cannot convince $V$ to reject the signature with probability greater than $\frac{1}{p(n,l)}$.
\end{theorem}
\begin{proof}
{\bf Completeness}:
Assume that $\hat{S}_m \notin SIG(m)$. We have,
\begin{equation*}
\begin{split}
&R_{1} = u_{i}a^{-1}(\hat{S}_m)^{2}x_{i}av_{i}^{-1} = a^{-1}u_{i}(\hat{S}_m)^{2}u_{i}^{-1}\beta a~~\mbox{ and}\\
&R_{2} = u_{i}b^{-1}(\hat{S}_m)^{2}x_{i}bv_{i}^{-1} = b^{-1}u_{i}(\hat{S}_m)^{2}u_{i}^{-1}\beta b.
\end{split}
\end{equation*}
Therefore,
\begin{equation*}
b^{-1}R_{1}b = a^{-1}R_{2}a = a^{-1}b^{-1}(u_{i}(\hat{S}_m)^{2}u_{i}^{-1}\beta)ba.
\end{equation*}
Hence the protocol is complete.\\
{\bf Soundness}:
Assume that $\hat{S}_m \in SIG(m)$. Let $R_{1}$ and $R_{2}$ be the responses given by $P^{*}$ in the protocol. Let if possible, $b^{-1}R_{1} b = a^{-1}R_{2}a.$  Then
\begin{equation*}
R_{2} = a(b^{-1}R_{1}b)a^{-1} =  a\beta a^{-1},\mbox{ where}\ \gamma = b^{-1}R_{1}b.
\end{equation*}
In the worst case, we may regard $\gamma$ as a known constant for $P$ when he tries to determine $R_{2}$. But then the ability to determine $R_{2}$ amounts to the establishment of an invalid signature, which contradicts Theorem 5.1 (soundness of the confirmation protocol). Hence the protocol is sound.
\end{proof}
\begin{remark}
For the ease of analysis, the disavowal protocol was given in a non zero-knowledge fashion. However, zero-knowledge versions of the disavowal protocol can also be constructed in a similar manner (see~\cite{tony}).
\end{remark}
\section{Group Signature Scheme 3}
In this section, we describe another group signature scheme. This scheme is given in the usual frame work of group signature schemes as described in~\cite{Pop}.  The security of the scheme is based on the hardnes of {\bf CSP, MSCSP} and {\bf MSBDP}. Here the recipient of the signature can easily verify the group identity of the signature. However, if a dispute occurs the group manager can open the signature and identify the signer.
\subsection{Setup}
The group manager $T$ chooses a secret braid $s \in_r LB_{n}(l)$, $k_1,k_2 \in_r RB_{n}(l)$,  and $\alpha \in_r B_{n}(l)$ and publishes $x = s^{-1}\alpha s$ as the public key of the group.
\subsection{Join}
Suppose now that a user $P$ wants to join the group. We assume that the communication between a group member and $T$ is secure, that is private and authentic.\par The following protocol is performed between the user $P$ and the Trusted Authority $T$.
\begin{enumerate}
\item $T \stackrel{(s, \alpha)}{\longrightarrow} P$.
\item $P$ chooses $u \in_r B_n(l)$ , $a \in_r LB_n(l)$ computes $v = u^{-1}\alpha u$, $w = a^{-1}u a$ and $P \stackrel{(v,w)}{\longrightarrow} T$.
\item $T$ computes $z_1 = k_{1}^{-1}w k_{1}$, $z_2 = k_{2}^{-1}w k_{2}$  and $T \stackrel{(z_1,z_2)}{\longrightarrow} P$.
\item $P$ computes $\beta_1 = a z_1 a^{-1}$ and  $\beta_2 = a z_2 a^{-1}$.
\end{enumerate}
Consequently, at the end of the protocol, $T$ creates a new entry in the group
 database with $v$ as the public key of the member $P$.
\subsection{Sign}
Let $m$ be the message which has to be signed. Suppose that the group member $P$ wants to sign $m$. He computes $S_{1} = s^{-1}ys$ and $S_{2} = s^{-1}\beta_{1}^{-1}y\beta_{2} s$, where $y = H(m)$. Signature is the pair $S_m = (S_{1}, S_{2})$.
\subsection{Verify}
A recipient of the signature after getting $S_{m}$, checks whether $S_{1}$ is conjugate to $y$ to check whether $S_m$ is a valid signature of $y$ or not.\par
To check the group identity of the signature, $V$ checks whether $S_{1}x$ is conjugate to $y\alpha$. If it holds, $V$ accepts $S_m$ as a signature from the group $G$.
\subsection{Open}
In case of a dispute, the group manager can identify the signer of the signature $\hat{S}_m = (\hat{S}_{1}, \hat{S}_{2})$ in the following way. He first computes $\hat{S}_{3} = k_1s\hat{S}_{2}s^{-1}k_{2}^{-1}$. Now he can find out whether $P$ is the signer by checking whether $\hat{S}_{3}v$ is conjugate to $k_{1}yk_{2}^{-1} \alpha$ or not. If it holds, the signature was made by $P$.
\subsection{Security Analysis}
In this section we will show that this group signature scheme satisfies some of the properties of for an ideal group signature.
\begin{enumerate}
\item{\bf Unforgeability}: Since to sign on behalf of the group,
one should know the secret key $s$, only group members can sign on
behalf of the group.  However, an attacker gets several pairs of
braids and its conjugates by $s$. Hence under the assumption that
multiple simultaneous conjugacy decomposition problem (MSCDP) is hard in braid
groups an attacker cannot get $s$ and the signature scheme stands
unforgeable.
\begin{remark}
We may make our scheme more secure by avoiding an attack on MSCDP in the following way : the group
manager chooses $s_1,s_2 \in_r LB_{n}(l)$ instead of $s \in_r
LB_{n}(l)$. He makes the group public key as $s_{2}^{-1}\alpha
s_1$. Now, given a message $m$ the signer computes the signature as
$S_m = (S_{1} = s_{1}^{-1}ys_2$, $S_{2} =
s_{1}\beta_{1}^{-1}y\beta_{2} s_2)$. The protocols for
verification and opening the signature can be rewritten in a
similar way.
\end{remark}
\item{\bf Unlinkability}: Let $m_1$ and $m_2$ be two messages signed by the group members. Let $y_1 = H(m_1)$ and  $y_2 = H(m_2)$. Let $S_{m_1} = (S_{1}^{1}, S_{2}^{1})$ and $S_{m_2} = (S_{1}^{2}, S_{2}^{2})$. Now, the problem of linking $S_{m_1}$ and $S_{m_2}$ reduces to deciding whether $S_{2}^{1}$ and $S_{2}^{2}$ are linked or not. Now, $S_{2}^{1} = s^{-1}\beta_{1}^{-1}y_1\beta_{2} s$ and $S_{2}^{2} = s^{-1}\beta_{1}^{-1}y_2\beta_{2} s$. Hence deciding whether $S_{2}^{1}$ and $S_{2}^{2}$ are linked or not reduces to checking whether the pairs $(S_{2}^{1},y_1)$ and $(S_{2}^{2},y_2)$ have the same factors or not. Now, this is infeasible by Assumption 3.1. Hence the signature scheme is unlinkable.
\item{\bf Anonymity}: Given a group signature, to identify the actual signer is computationally hard to do for everyone but the group manager. Consider a  signature on $m$ by $P$.  Let $S_{m} = (S_{1}, S_{2})$. Now $S_{2} = s^{-1}(k_{1}^{-1}u^{-1}k_{1})y(k_{2}^{-1}uk_{2})s$ and in order to show that the signature belongs to $P$, a group member has to prove that  $(k_{1}s\hat{S}_{2}s^{-1}k_{2}^{-1})v$ is conjugate to $k_{1}yk_{2}^{-1} \alpha$. There is no apparent way of proving the identity of the signer other than by getting the private keys of the signer or that of the Trusted Authority. But any group member can compute $sS_{2}s^{-1} = (k_{1}^{-1}u^{-1}k_{1})y(k_{2}^{-1}uk_{2})$. Now, the only way for a group member $\hat{P}$ with secret key $\hat{v}$ to find out the identity of the signer is to get the value of $k_1$ and $k_2$ from $k_{1}^{-1}\hat{v}k_{1}$ and $k_{2}^{-1}\hat{v}k_{2}$ which he obtained from the group manger. But this amounts to solving a conjugacy search problem and by assumption the conjugacy search problem is hard. Hence the signature scheme is anonymous.
\item{\bf Exculpability }: The group manager does not get any information about a group member's secret key $u$ as well as signing keys $k_{1}^{-1}uk_{1}$ and $k_{2}^{-1}uk_{2}$. The values of $u$ as well $k_{1}^{-1}uk_{1}$ and $k_{2}^{-1}uk_{2}$ are computationally hidden from the group manager because of the protocols involved in the Join session of the member $P$. Hence the group manager cannot sign on behalf of a group member.  Similarly, any group member cannot sign on behalf of any other member. Hence exculpability holds.
\item{\bf Traceability}: Assume that the signature $S_m = (S_{1}, S_{2})$ on the message $m$ was made by $P$. Now the group manager can compute
\begin{eqnarray*}
S_{3}v &=& (k_1sS_{2}s^{-1}k_{2}^{-1})v = (k_1s(s^{-1}\beta_{1}^{-1}y\beta_{2} s)s^{-1}k_{2}^{-1})(u^{-1}\alpha u)\\
&=&(k_1\beta_{1}^{-1}y\beta_{2}k_{2}^{-1})(u^{-1}\alpha u)=(k_1(k_{1}^{-1}u^{-1}k_1)y(k_{2}^{-1}uk_{2})k_{2}^{-1})(u^{-1}\alpha u)\\
&=&u^{-1}k_1yk_{2}^{-1}\alpha u.
\end{eqnarray*}
Hence $S_{3}v$ is conjugate to $k_{1}yk_{2}^{-1} \alpha$. Thus, the group manager can open any valid group signature and identify the actual signer. Hence the signature is traceable.
\end{enumerate}
\section{Concluding Remarks}
In this paper, we constructed three group signature schemes
based on some hard problems in braid groups. Our schemes are the
first in this direction using braid groups. It is open to use other hard problems in braid groups for designing more
group signature schemes and other cryptographic protocols.\par
The first signature scheme has the property that
its security is entirely depending on the root problem. This is
the only cryptographic scheme on braid groups whose security is
solely depending on the root problem. Root problem is believed to
be harder than the conjugacy and decomposition problems. Hence
we may believe that this scheme is the most secure one. The
second scheme  combines the notion of undeniable signatures with group signatures. Our third scheme is set in the usual frame work of group signatures.\par
 The problem of checking the equality of factors in a 3-factor decomposition of
two given braids with the middle factors known is employed in
Assumption 3.1. We leave this assumption as well as Assumption 3.2 for further investigation. The first step in the investigation of the second assumption may be  to the estimate of number of conjugates of a random element which are equal. Numerical experiments might throw some light on these assumptions.\par 
 The birth of braid cryptography has simulated the search for other
exotic mathematical structures for doing public-key cryptography.
 People have started looking at other nonabelian
groups~\cite{stick},~\cite{stein},~\cite{paeng},~\cite{gri} and
combinatorial groups~\cite{vlad},~\cite{vlsh} for building
public-key cryptosystems. Although, we have described our schemes in the frame work of braid groups, these protocols can be carried over to many other nonabelian groups with slight modifications. Further, one can modify these protocols to other variations of group signatures like, the ring signatures and
undeniable group signatures discussed in Section 1. Hence, we hope that
this study will motivate further research on digital signatures
based on nonabelian groups and combinatorial groups.

\end{document}